\documentclass[]{elsart}
\usepackage{graphicx,amssymb,psfig}
\journal{New Astronomy}
\def\astrobj#1{#1}
\begin{document}
\begin{frontmatter}

\title{Photometric study of \astrobj{HD\,155555}C\\ in the $\beta$ Pictoris Association}
\author[INAF]{Sergio Messina\corauthref{cor}},
\corauth[cor]{Corresponding author.}
\ead{sergio.messina@oact.inaf.it}
\author[YCO]{Mervyn Millward},
\ead{mervyn.millward@yorkcreek.net}
\author[EU]{David H. Bradstreet}
\ead{dbradstr@eastern.edu}
\address[INAF]{INAF- Catania Astrophysical Observatory, via S.Sofia, 78 I-95123 Catania, Italy} 
\address[YCO]{York Creek Observatory, Georgetown, Tasmania}
\address[EU]{Eastern University, St. Davids, PA, USA}
 
\begin{abstract} 
We are carrying out a series of photometric monitoring to measure the rotation periods of members in the young \astrobj{$\beta$ Pictoris Association}, as part of the RACE-OC project (Rotation and ACtivity Evolution in Open Clusters). In this paper, we present the results for 
\astrobj{HD\,155555C} which is believed to be physically associated to the spectroscopic binary \astrobj{V824 Ara} (HD155555) and thus constituting a triple system. We collected B, V, and R-band photometric data timeseries and discovered from periodogram analysis the rotation period P = 4.43\,d. Combined with stellar radius and projected rotational velocity, we find this star almost equator-on with an inclination $i$ $\simeq$ 90$^{\circ}$. The rotational properties of HD155555C  fit well into the period distribution of other $\beta$ Pic members, giving further support to the suggested  membership to the association and to its physical association to \astrobj{V824 Ara}. A comparison with Pre-Main-Sequence isochrones from various models allows us to estimate an age of 20$\pm$15 Myr for this triple system.
\end{abstract}
\end{frontmatter}

\section{Introduction}
We are carrying out a systematic investigation aimed at determining the rotation periods and the photospheric activity levels of all confirmed and candidate members of the young $\beta$ Pictoris stellar Association (Zuckerman et al. 2001).
This investigation is part of the RACE-OC project (Rotation and Activity Evolution in Open Clusters; Messina \cite{Messina07}). In an earlier study by Messina et al. \cite{Messina10}, we focused on only confirmed members and derived the rotation periods for a total of 27 of them either from the literature or from our analysis of  photometric timeseries in the  ASAS (All Sky Automated Survey; Pojmanski \cite{Pojmanski97}) and SuperWASP (Wide Angle Search for Planets; Butters et al. \cite{Butters10}) archives. However,
a few members have remained with  unknown or uncertain periods for a number of reasons:  they were not in the archives; the low-photometric precision in combination with low variability level prevented us from rotation period detection;  they were in spatially unresolved  multiple systems. Meanwhile, new candidate members have been proposed, e.g., by  Lepine et al. \cite{Lepine09}, Kiss et al. \cite{Kiss11}, Shkolnik et al.  \cite{Shkolnik12}, Schlieder et al. \cite{Schlieder12}, Malo et al. \cite{Malo14}.
In order to have a complete rotational characterization of the association, we have undertaken a number of collaborations.
In the present paper, we present the results of the monitoring project carried out at the York Creek Observatory  on \astrobj{HD\,155555C}, one of the first proposed members to this Association.\\
\indent
We focus our investigation on the pre-main-sequence triple system \astrobj{HD\,155555} consisting of a SB2 spectroscopic binary, \astrobj{HD\,155555AB}, and a nearby single dM component, \astrobj{HD\,155555C}. 
\astrobj{HD\,155555AB} has been identified as member of the young stellar Association $\beta$ Pictoris by Zuckerman et al. \cite{Zuckerman01}. The nearby dM star has been suggested to be physically connected to the SB2  by Pasquini et al. \cite{Pasquini89} and Pasquini et al. \cite{Pasquini91}, and therefore to be a member of the same Association. However, no parallax of the latter is measured, and membership to this stellar system as well as to the Association are only inferred, though significantly supported by a number of evidences. Our rotational investigation aims to add additional information to get a better characterization of this dM component.\\
\indent
For instance, this system, like other multiple systems, see, e.g., Messina et al. \cite{Messina14}, is particularly interesting since a few parameters are equal, like age, initial chemical composition, and mass for some systems. These commonalities allow us to explore the dependence of their rotational evolution on other parameters, such as initial angular velocity and environment properties (presence of disc or nearby stellar/planetary companions), shedding light on the causes of dispersion observed in the rotation period distributions of all young clusters/associations.

\section{Literature information}

\astrobj{HD\,155555} is a triple stellar system in the southern constellation of Ara  at a distance d = 31.4$\pm$0.5 pc  (van Leeuwen \cite{vanLeeuwen07}). \rm This system consists of a double-line spectroscopic binary \astrobj{HD\,155555AB} and of a fainter M dwarf companion \astrobj{HD\,155555C} at an angular distance of about 33$^{\prime\prime}$,  which corresponds to about 1040 AU. \rm \\
\indent
The spectroscopic binary \astrobj{HD\,155555AB} was discovered by Bennet et al. \cite{Bennett67} to be composed of  G5IV + K0IV stars that rotate with an orbital period P = 1.687\,d.  First evidence of photometric variability of this SB2 was found by Eggen \cite{Eggen78} after comparing his photometry collected in 1978 (V = 6.83, B$-$V = 0.835, U$-$B = 0.29, R = 6.46, and R$-$I = 0.325 mag) with the earlier photometry collected by Stoy \cite{Stoy63} (V = 6.67 and B$-$V = 0.80 mag), even though the latter also included the dM companion, owing to the large aperture of 44$^{\prime\prime}$ used to collect the photoelectric photometry.  Subsequent long-term photometric monitoring by Cutispoto (see, e.g., Cutispoto 1998 and references therein) revealed robust evidence of the existence of short- and long-term photometric variability. \astrobj{HD\,155555AB} has the variable's name \astrobj{V824 Ara} in the Combined General Catalogue of Variable Stars \cite{Kholopov98}. The most recent determination of orbital and physical properties,   and first magnetic maps \rm of both components were provided by the spectro-polarimetric study of Dunstone et al. \cite{Dunstone98}.   Most recent surface temperature maps have been obtained with an improved Doppler imaging technique by Kristovics et al. \cite{Kriskovics13}. \rm 
The membership of HD155555AB to the \astrobj{$\beta$ Pictoris} moving group was first suggested by Zuckerman et al. \cite{Zuckerman01} based on distance, UVW
velocity components, high $v\sin{i}$, and L$_{\rm X}$/ L$_{\rm bol}$ ratio.  Also the photometric variability observed by Eggen (1978) and  Cutispoto (1998) are in agreement with the youth of the system. \rm\\
\indent
The present investigation is focused on the fainter \astrobj{HD\,155555C} component (also named  \astrobj{LDS\, 587B}) which is a M3.5 dwarf  (Pasquini et al. 1991). Eggen \cite{Eggen78} reported the following magnitudes and colors: V = 12.82, B$-$V = 1.54, U$-$B = 1.05, R = 11.40, and R$-$I = 1.205 mag, noticing both UV and IR flux excess with respect to standard colors of main sequence stars. This component is a strong X-ray source (\astrobj{EXO\,171224-6653.9}), which was a serendipitous discovery by Barstow et al. \cite{Barstow87} during EXOSAT observations of \astrobj{HD\,155555AB} (\astrobj{EXO\,171217-6653.8}). Pasquini et al. \cite{Pasquini89} identified this X-ray source with \astrobj{HD\,155555C}, and subsequent  spectroscopic investigation (Pasquini et al. 1991) showed high levels of magnetic activity, the HeI, NaI D2 and D1 lines all being in emission. The H$\alpha$ line was also found to be in emission and variable from night to night with  the equivalent widths \rm ranging from 6.2 to 9.7 \AA. Pasquini et al. \cite{Pasquini91}  could measure only an upper value  ($\le 3$ km\,s$^{-1}$) of the projected rotational velocity. Their study  concluded that HD155555\,AB and HD155555\,C are physically associated and are likely PMS stars. 
They noticed that radial velocities are similar (V$_r$ = 2.74\,km\,s$^{-1}$ for \astrobj{HD\,155555AB} compared to  V$_r$ = 3.5\,km\,s$^{-1}$ for \astrobj{HD\,155555C}. 
Torres et al. \cite{Torres06} also find that the UVW velocity components and radial velocity of HD155555C   are similar to those of HD155555AB, suggesting their physical association.\\
\indent
The physical association is also supported by proper motion studies by Martin et al. \cite{Martin95}, who find the angular separation changed from 33$^{\prime\prime}$ (about in 1929) to 32.6$^{\prime\prime}$ in 1994, suggesting similar proper motions.  Similarly, the Washington Visual Double Star Catalog \cite{Mason01} reports a small variation from 33$^{\prime\prime}$ in 1920 to 34$^{\prime\prime}$ in 2000.
In the literature, two other measurements of projected rotational velocity are available from Torres et al. \cite{Torres06}  $v \sin{i}=6.0\pm1.2$\,km\,s$^{-1}$ and from Weise et al. \cite{Weise10}  $v \sin{i}=8.8\pm0.9$\,km\,s$^{-1}$.

 \begin{figure}
\begin{minipage}{10cm}
\centerline{
\includegraphics[width=90mm,height=120mm,angle=90,trim= 0 0 0 0]{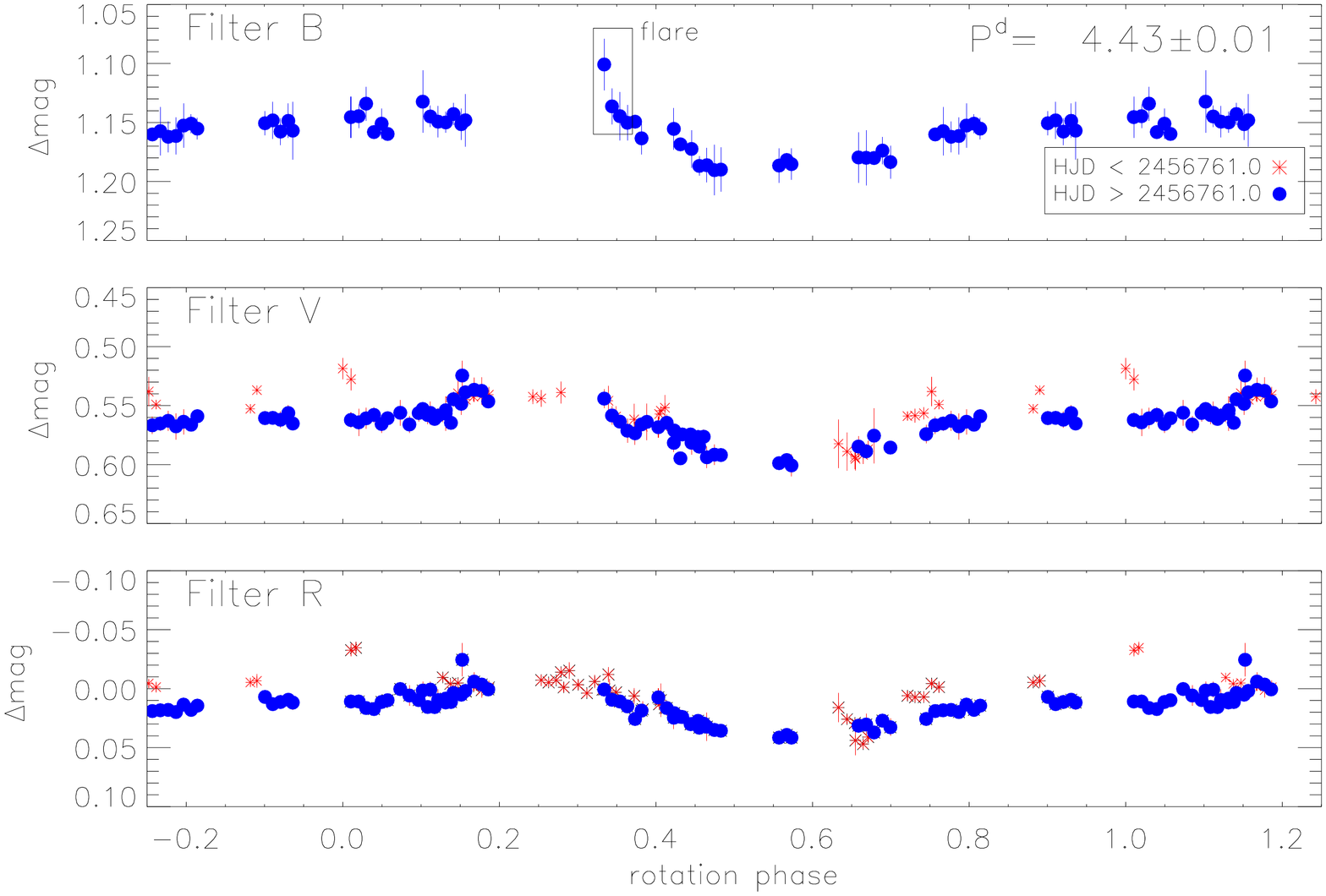} \\
}
\end{minipage}
\caption{\label{lightcurve}Light curves of HD155555C phased with the rotation period  in three filters. Blue bullets denote observations collected after HJD 2456761 to better
show the light curve evolution. The rectangle denotes a possible flare event.}
\vspace{0cm}
\end{figure}

 \begin{figure}
\begin{minipage}{10cm}
\centerline{
\includegraphics[width=90mm,height=120mm,angle=90,trim= 0 0 0 0]{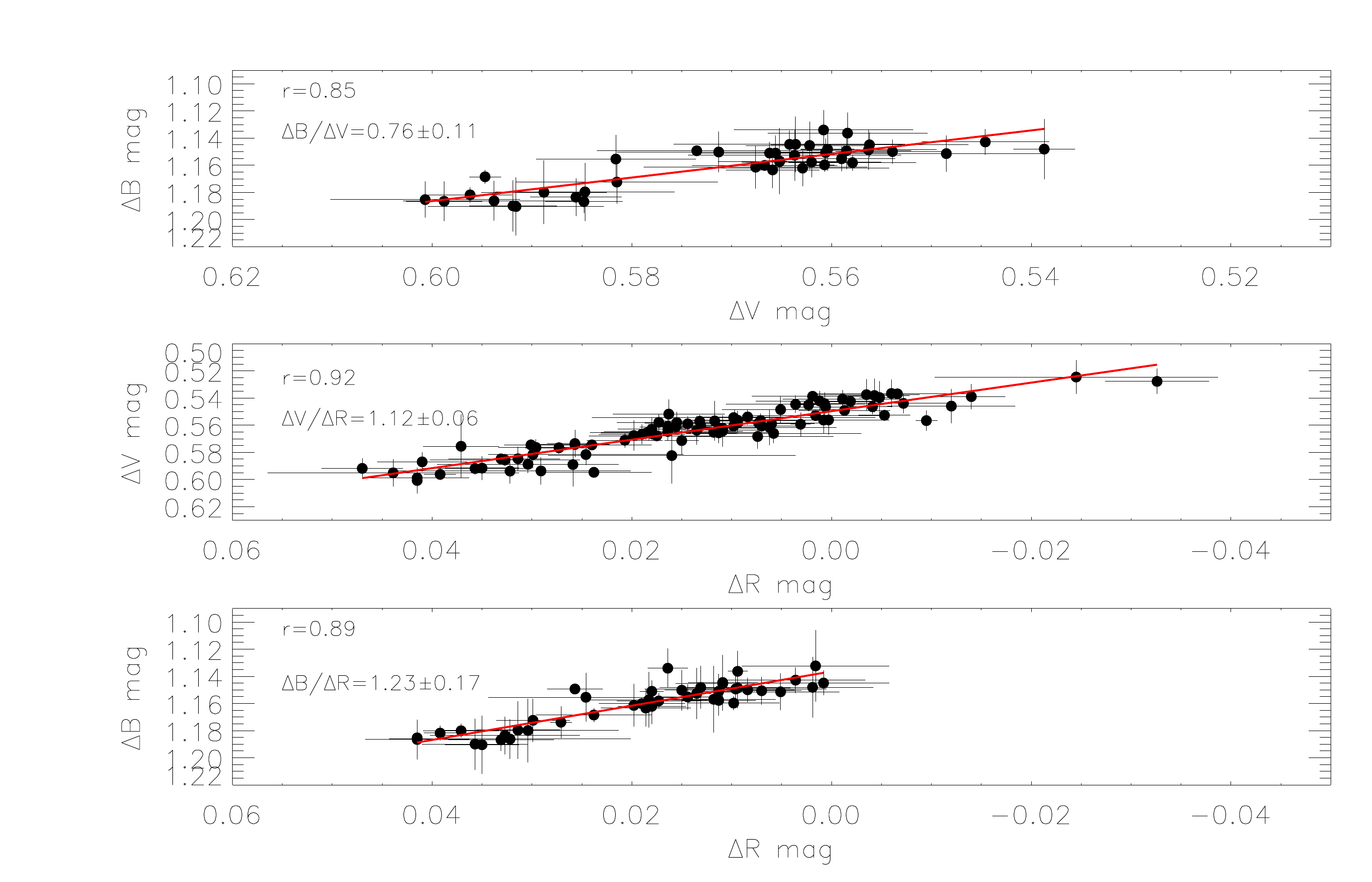} \\
}
\end{minipage}
\caption{\label{correl} Correlations among magnitude variations in the three filters. Solid lines are  weighted \rm linear fits, whereas labels indicates the linear correlation coefficient (r) and the fit slope with its uncertainty. }
\vspace{0cm}
\end{figure}  

 \begin{figure}
\begin{minipage}{18cm}
\includegraphics[width=50mm,height=60mm,angle=90,trim= 0 0 0 0]{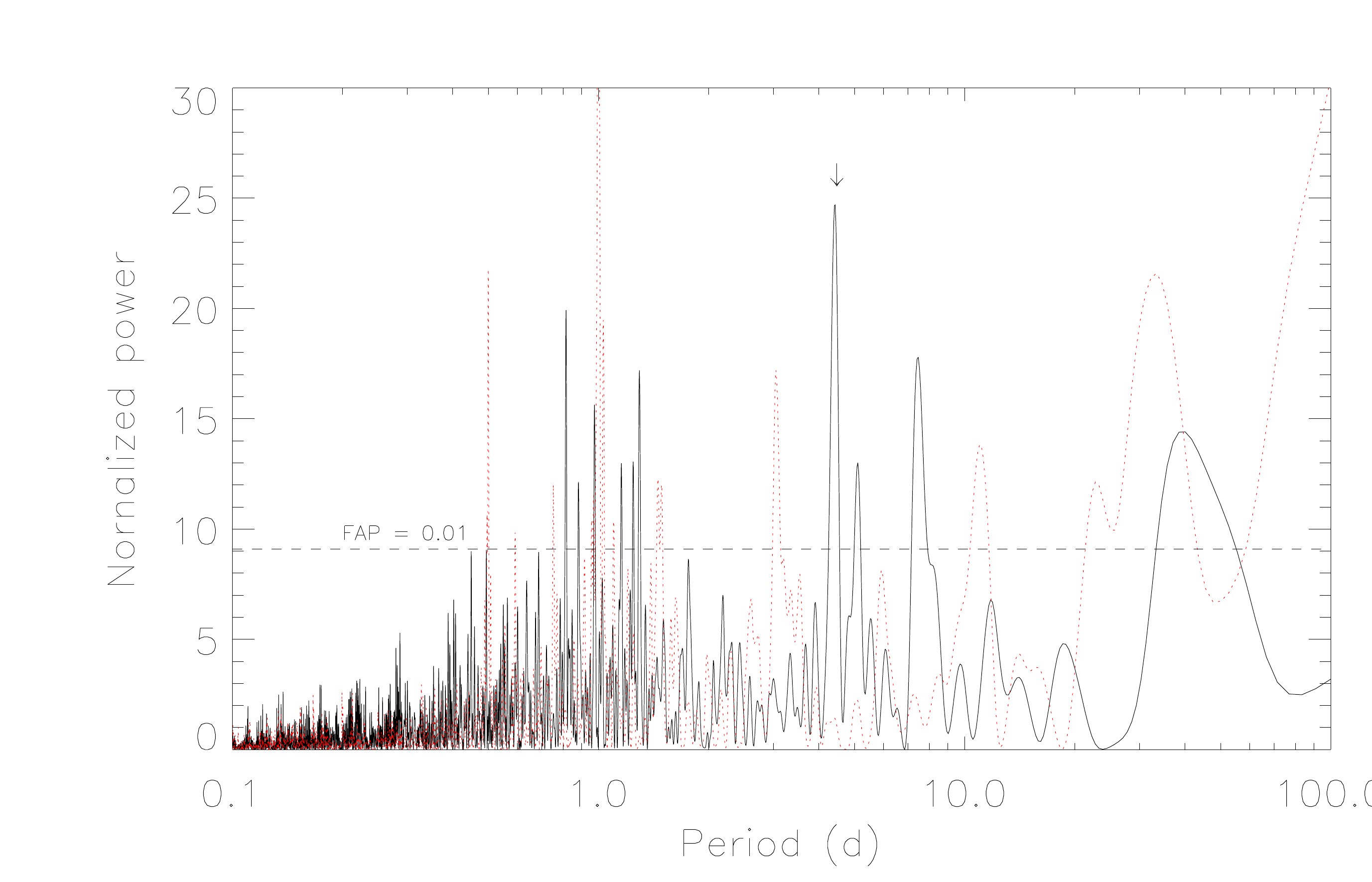} 
\includegraphics[width=50mm,height=60mm,angle=90,trim= 0 0 0 0]{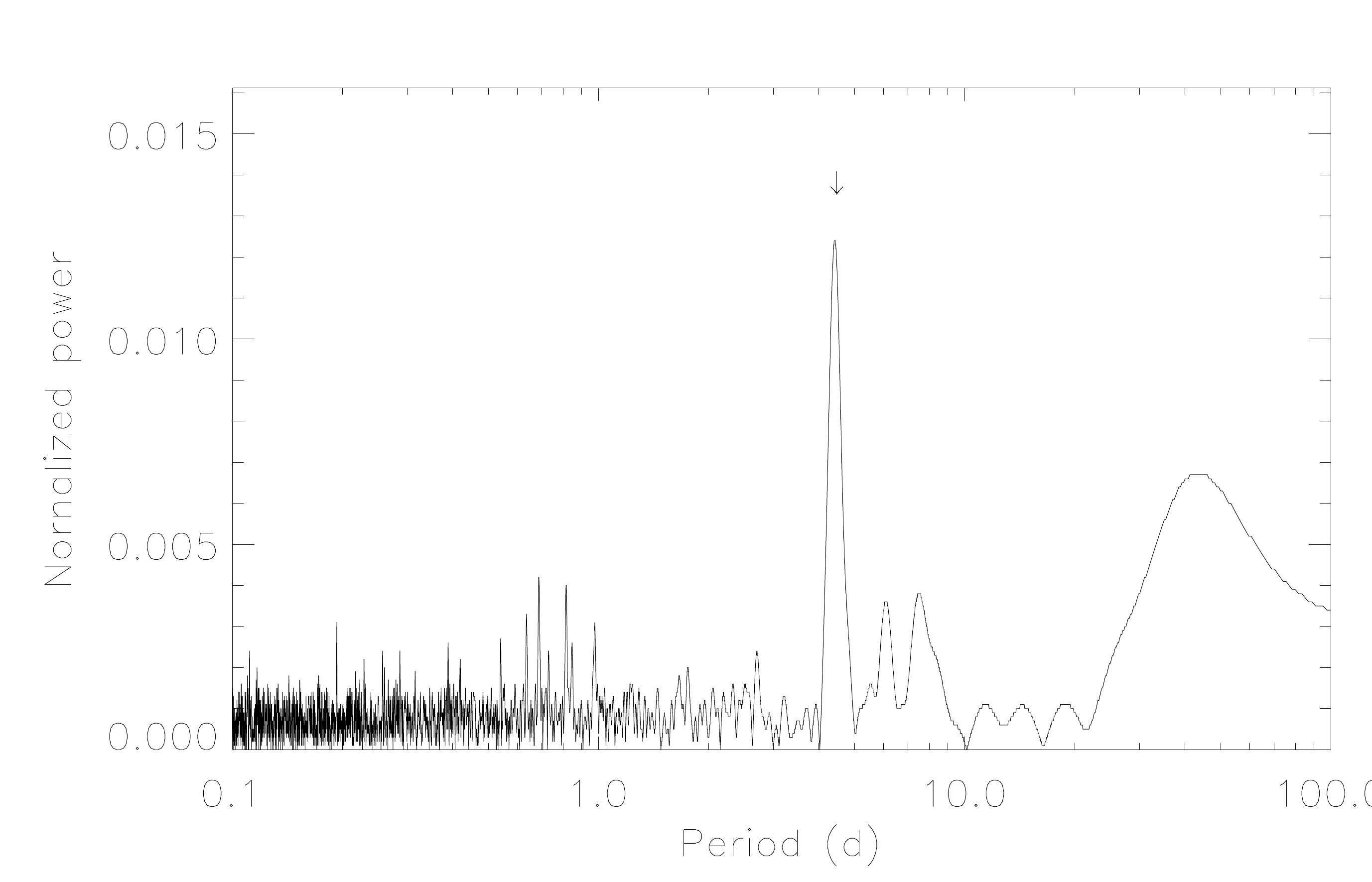} 
\end{minipage}
\caption{\label{periodogram}\it left panel\rm: Lomb-Scargle periodogram of HD155555C (solid line) with indication of the power peak corresponding to the rotation period P = 4.43\,d. The dotted red line indicates the window spectral function, whereas the horizontal dashed line the power level corresponding to a FAP = 0.01. \it right panel\rm: Clean periodogram.}
\end{figure}  

 \begin{figure}
\begin{minipage}{18cm}
\includegraphics[width=100mm,height=130mm,angle=90,trim= 0 0 0 0]{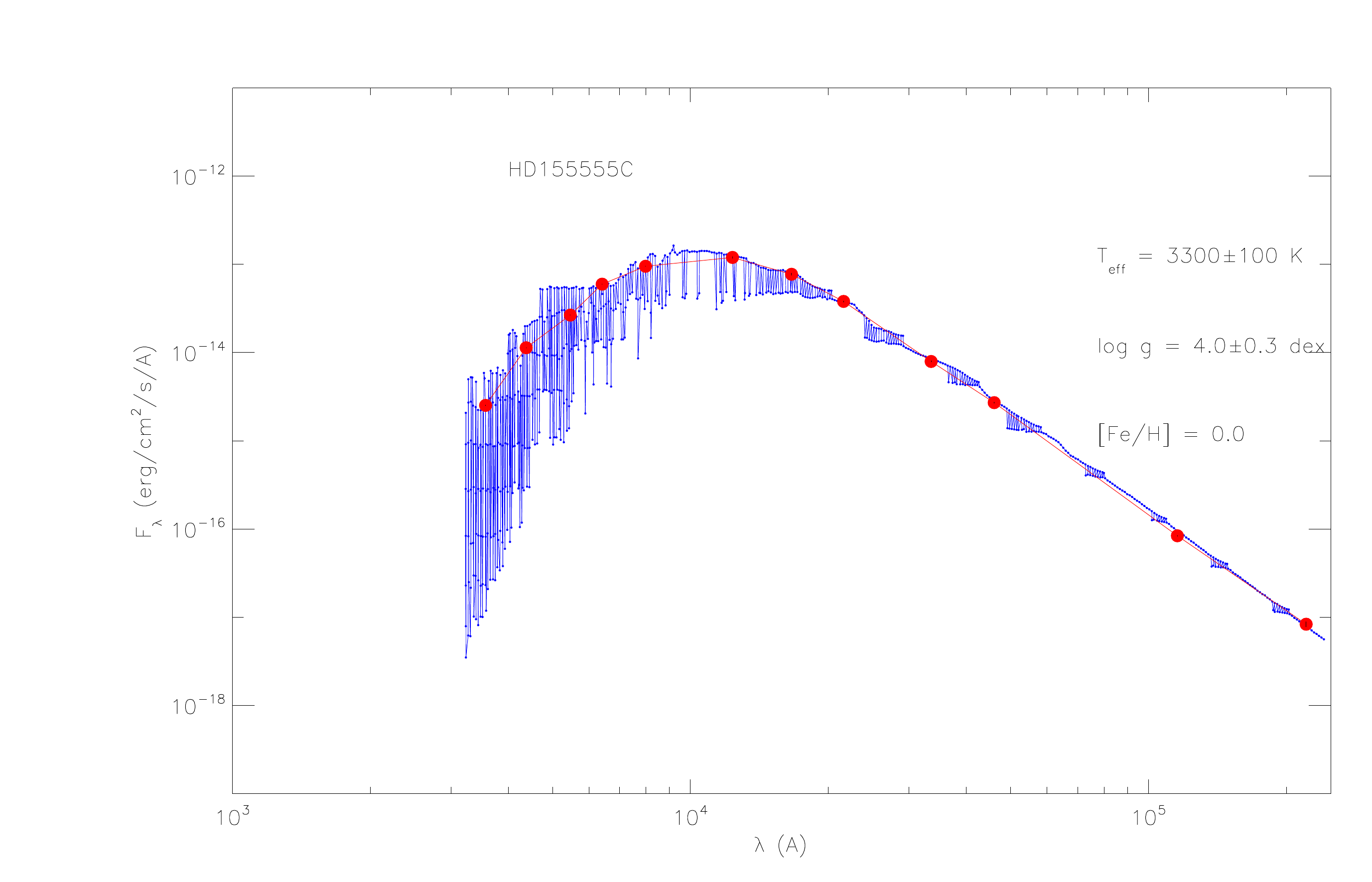} 
\end{minipage}
\caption{\label{sed}Spectral Energy Distribution  of HD155555C.}
\vspace{0cm}
\end{figure}  

This system has been searched for possible planetary companions using the NACO adaptive optics system at the  Very Large Telescope (VLT) \cite{Masciadri05}, and the MMT telescope \cite{Biller07}. It was included in the SIM PlanetQuest Key Project \cite{Tanner07}, and more recently in the Gemini/NICI planet-finding campaign \cite{Biller13}. None of these investigations have so far detected  the presence of any additional companion. 

\section{Photometric observations}
\astrobj{HD\,155555C} was observed in 2014, from Feb 28 to May 10, for a total of 21 nights. Observations were collected at the York Creek Observatory (41$^{\circ}$\,06$^{\prime}$\,06$^{\prime\prime}$S; 146$^{\circ}$\,50$^{\prime}$\,33$^{\prime\prime}$E, Georgetown, Tasmania) using a f/10 25cm Takahashi Mewlon reflector, equipped with a QSI 683ws-8 camera, and BVR standard Johnson-Cousins filters. The telescope has a FoV of 24.5$^{\prime}$$\times$18.5$^{\prime}$, and the plate scale is 0.44$^{\prime\prime}$/pixels. On each night observations were collected from 3 to 6 consecutive hours, using 120-s integrations in each filter. A total of 360 frames in B, 359 in V, and 400 in R filter were obtained (see Table\,\ref{log}).

After bias subtraction and flat fielding, we extracted  the magnitude timeseries for a total of 30 stars. These series were cleaned by applying a 3$\sigma$ threshold to remove outliers. 
The differential photometry was performed using the technique of
ensemble photometry (Gilliland et al. 1988; Bailer-Jones et al. 2001; Everett et al. 2001). The differential magnitude of the target star
is computed with respect to the average magnitude of a large number of non-variable reference stars.
In this averaging process,
the uncertainties of the ensemble star magnitudes due to statistical
fluctuations as well as short-term small incoherent variations
will cancel each other. The uncertainty in the magnitude of the artificial
comparison will therefore be smaller than the uncertainty
on the magnitude of a single star and this, in turn, will produce
less noisy light curves. 

We started with 30 sufficiently bright stars, which were isolated, 
distributed all over the frame but not found to be
close to the CCD edges, and common to all the frames. Having selected 
the ensemble, we 
computed (i) the magnitude of the artificial comparison
star by averaging the instrumental magnitudes of ensemble stars, (ii)
time-series differential magnitudes for each star of the ensemble and
(iii) mean, median and the standard deviation ($\sigma$) associated with
each time series. 
While constructing the final ensemble, we excluded those ensemble stars whose
standard deviation was larger than the threshold $\sigma$ value that we fixed to 0.010 mag.
After a number of iterations, we were left with six stars whose standard deviations were smaller than 0.010 mag. In Table\,\ref{comp} we list the 2MASS designation \cite{Cutri03} comparison stars, the nearby optical sources from USNO catalogue \cite{Monet13}, the coordinates and the J-band magnitude. These stars were used to build the ensemble comparison star and to obtain the differential magnitude time series of \astrobj{HD\,155555}C in B, V, and R filters.\\
 In Fig.\,\ref{lightcurve}, we plot the B, V, and R light curves phased with the P = 4.43\,d rotation period (see Sect.\,5). The average photometric uncertainties associated to each magnitude are 0.014, 0.007, and 0.005 mag in the B, V, and R filters, respectively. \rm
 \begin{table*}
 \caption{\label{log} Log of observations.}
 \begin{tabular}{cccc}
 \hline
Observation & \multicolumn{3}{c}{\# frames}\\
Date   & B & V & R \\
\hline
28/02/2014	& 8& 10& 8\\
02/03/2014	& 10& 10& 43\\
19/03/2014	& 4& 4& 34\\
24/03/2014	& 3& 3& 3\\
25/03/2014	& 19& 16& 17\\
03/04/2014	& 3& 3& 3\\
04/04/2014	& 5& 6& 5\\
05/04/2014	& 20& 24& 24\\
06/04/2014	& 22& 20& 14\\
12/04/2014	& 22& 22& 19\\
14/04/2014	& 13& 15& 15\\
24/04/2014	& 32& 32& 29\\
27/04/2014	& 17& 17& 18\\
30/04/2014	& 33& 33& 33\\
03/05/2014	& 24& 26& 26\\
04/05/2014	& 20& 18& 17\\
05/05/2014	& 18& 18& 17\\
06/05/2014	& 27& 26& 27\\
07/05/2014	& 25& 24& 19\\
08/05/2014	& 10& 10& 9\\
10/05/2014	& 25& 22& 20\\
 \hline
  \end{tabular}
  \end{table*}

 \begin{table*}
 \caption{\label{comp} Comparison stars.}
 \begin{tabular}{ccccc}
 \hline
2MASS & USNO-B1.0 & RA  & DEC & J mag\\
\hline
\astrobj{2MASS J17174644-6656455} & 09064-02437   & 17 17 46.44    & $-$66 56 45.60    & 8.696 \\
\astrobj{2MASS J17174439-6657287} & 0230-0693233 & 17 17 44.37  & $-$66 57 28.84  &11.285\\
\astrobj{2MASS J17175417-6657550} & 0230-0693339 & 17 17 54.17  & $-$66 57 55.06  & 12.085\\
\astrobj{2MASS J17172719-6700468} & 0229-0763917 & 17 17 27.19  & $-$67 00 46.86 &11.457\\
\astrobj{2MASS J17172092-6701280} & 0229-0763856 & 17 17 20.93  & $-$67 01 28.08 & 9.411	\\
\astrobj{2MASS J17170984-6700057} & 0229-0763745 & 17 17 09.84  & $-$67 00 05.77 &12.383	\\
\hline
  \end{tabular}
  \end{table*}

\section{Analysis}
As shown in Fig.\,\ref{correl}, the photometric timeseries show that \astrobj{HD\,155555C} varies in all B, V, and R bands. The magnitude variations are strongly correlated to each other. The Pearson linear correlation coefficients are all positive, r$_{\rm BV}$ = 0.85, r$_{\rm VR}$ = 0.92, and r$_{\rm BR}$ = 0.89, and all with a significance level $>$ 99.9\%.   Weighted \rm linear fits to the magnitude variations (solid lines in Fig.\,\ref{correl}) give the following slopes  $\Delta$B/$\Delta$V = 0.76$\pm$0.11, $\Delta$V/$\Delta$R = 1.12$\pm$0.06, and $\Delta$B/$\Delta$R = 1.23$\pm$0.17. \rm
 This is the behavior typically exhibited,
according to Messina \cite{Messina08}, by \it color-correlated \rm and \it reddening  stars, \rm whose color variations arise from cool spots. \rm  Such temperature inhomogeneities generally give rise to light curves whose amplitudes (peak-to-peak) increase towards shorter wavelengths, that is the B light curve has amplitude larger than the V light curve that is larger than the R light
curve. In our specific case, we note that whereas V (whose amplitude is $\Delta$V = 0.07 mag) and R (whose amplitude is $\Delta$R = 0.05 mag) curves follow this trend, the B-band light curve has amplitude comparable, if not slightly smaller ($\Delta$B = 0.06 mag), than V ($\Delta$B/$\Delta$V $<$ 1).  Unfortunately, our B-band observations could not cover the rotation phases around the lightcurve maximum (phases 0.15--0.35). Therefore, the true amplitude of variation in the B band may be larger than measured and  certainly further observations will address this aspect. \rm The V and R light curves, for which we have the longest timeseries,  exhibit evidence of a significant variation, the amplitude has decreased and the phase of light maximum migrated by $\Delta \phi$ = 0.2, as likely consequence of the active regions growth and decay (ARGD), which are responsible for the observed variability.

\section{Rotation period search}
We used the Lomb-Scargle \cite{Scargle82} and Clean \cite{Roberts87} periodogram analyses to search for significant periodicities in the HD155555C magnitude timeseries
related to stellar rotation period. The results of our analysis are plotted in Fig.\,\ref{periodogram}. In the left panel we plot the normalized Lomb-Scargle periodogram (solid line) with overplotted the spectral window function related to the data sampling (red dotted line). The horizontal dashed line indicates the power level corresponding to a False Alarm Probability (FAP) of 1\%, that is the probability that a power peak of that height simply arises from Gaussian noise in the data. The FAP was estimated using a Monte-Carlo method, i.e., by generating 1000 artificial light curves obtained from the real one, keeping the date but scrambling the magnitude values. In the right panel we plot the Clean periodogram where the power peak arising from the light rotational modulation dominates, whereas all secondary peaks, arising from the aliasing, are effectively removed. Our analysis finds that the stellar rotation period is P = 4.43$\pm$0.05\,d. The uncertainty of the rotation period is computed following the prescription of Lamm et al. \cite{Lamm04}.

\section{SED analysis}
We used the available optical, near-IR, and IR photometry to build the observed spectral energy distribution (SED). The UBVRI magnitudes from Eggen \cite{Eggen78}, JHK magnitudes from 2MASS \cite{Cutri03}, and W1-W4 magnitudes from WISE \cite{Cutri13} are listed in Table\,\ref{photometry}. 
The SED was fitted with a grid of theoretical spectra from the NextGen Model \cite{Allard12} and the best fit is obtained with a model of   T$_{\rm eff}$ = 3300$\pm$100\,K, log g = 4.0$\pm$0.3 dex \rm and metallicity [Fe/H] = 0.0  (see Fig.\ref{sed}). We note that no evidence of IR excess was found.

\section{Target parameters}
\begin{figure}
\begin{minipage}{10cm}
\includegraphics[width=50mm,height=80mm,angle=90,trim= 0 0 0 0]{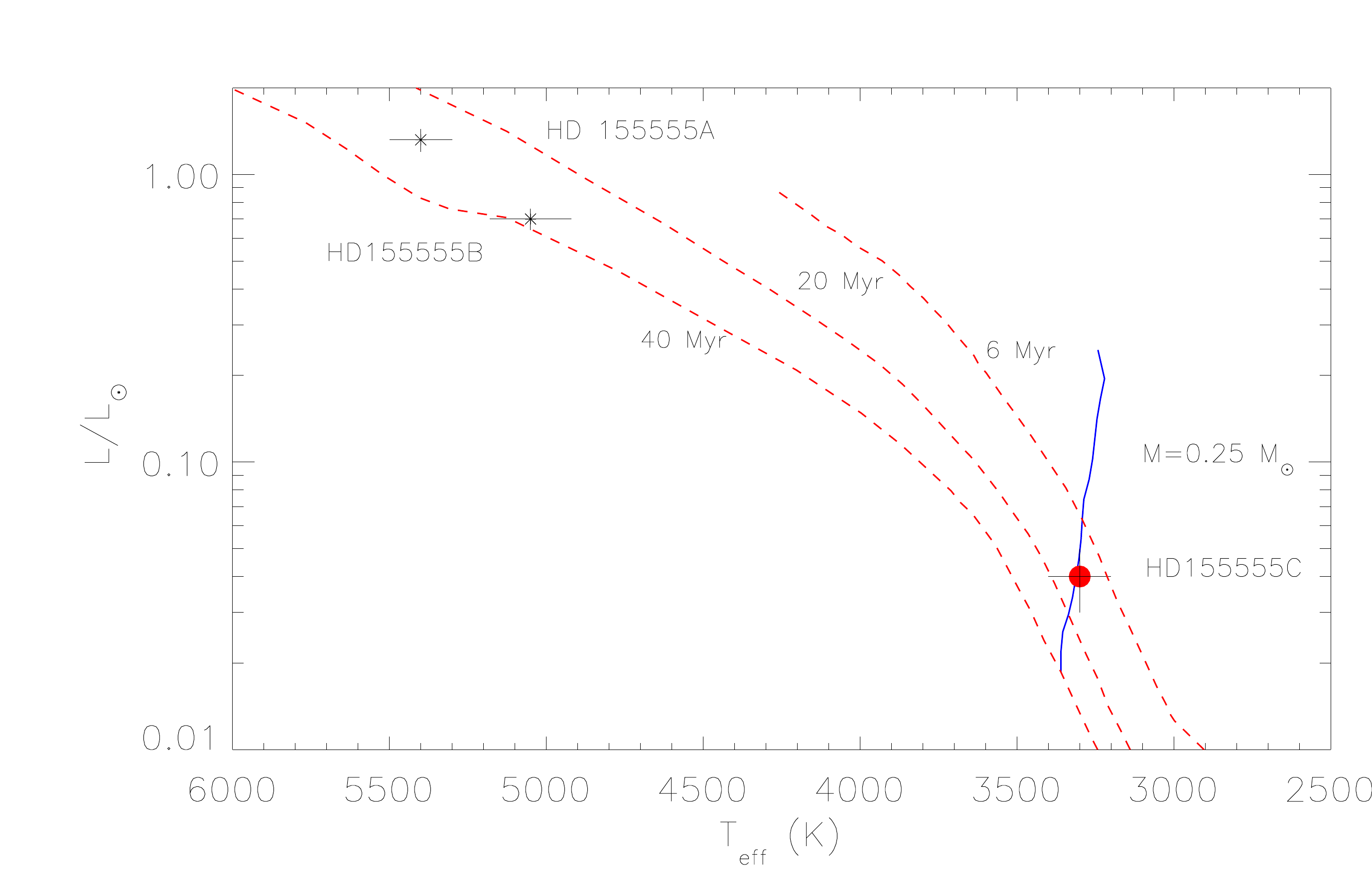} 
\includegraphics[width=50mm,height=80mm,angle=90,trim= 0 0 0 0]{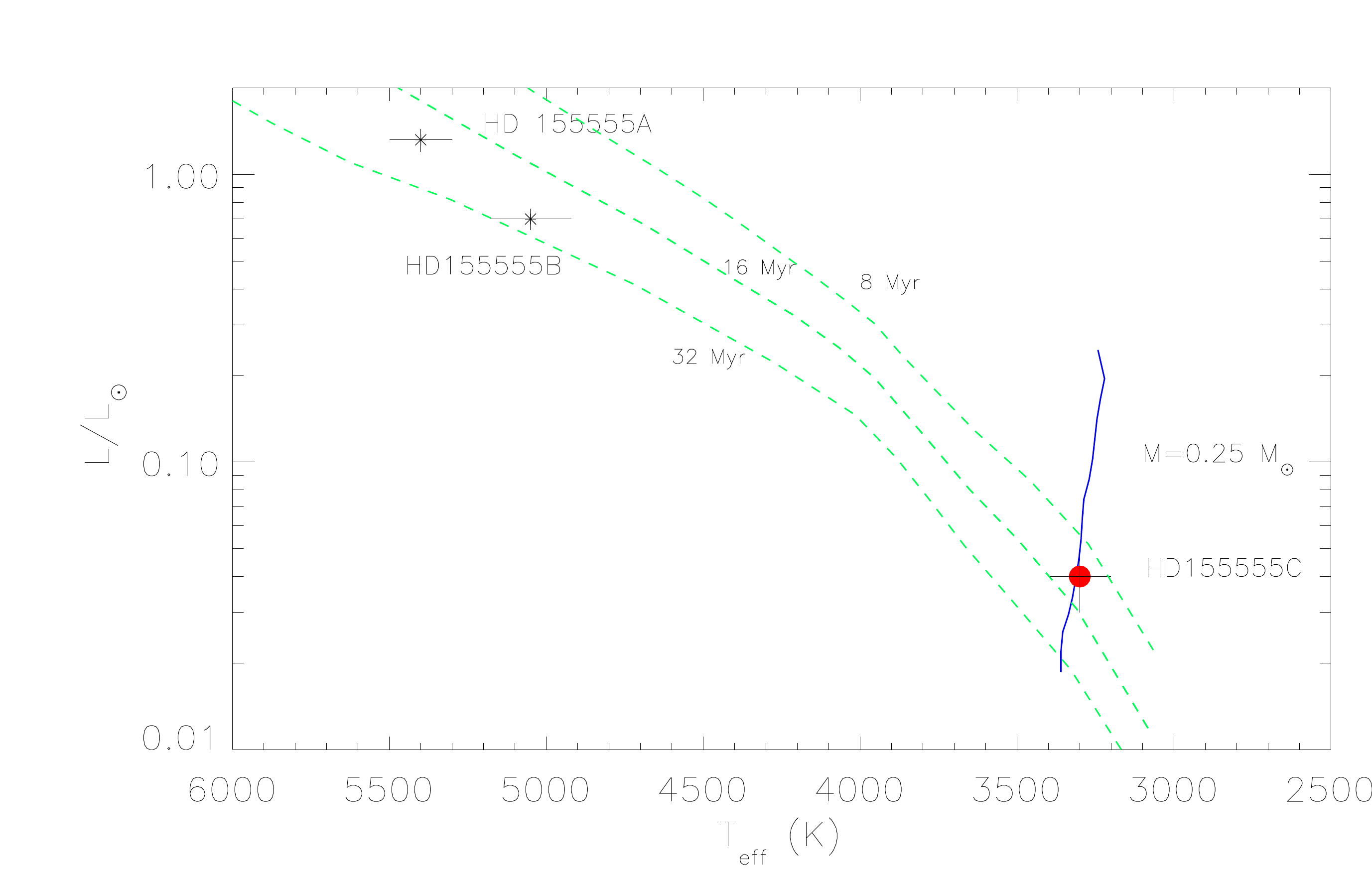} 
\includegraphics[width=50mm,height=80mm,angle=90,trim= 0 0 0 0]{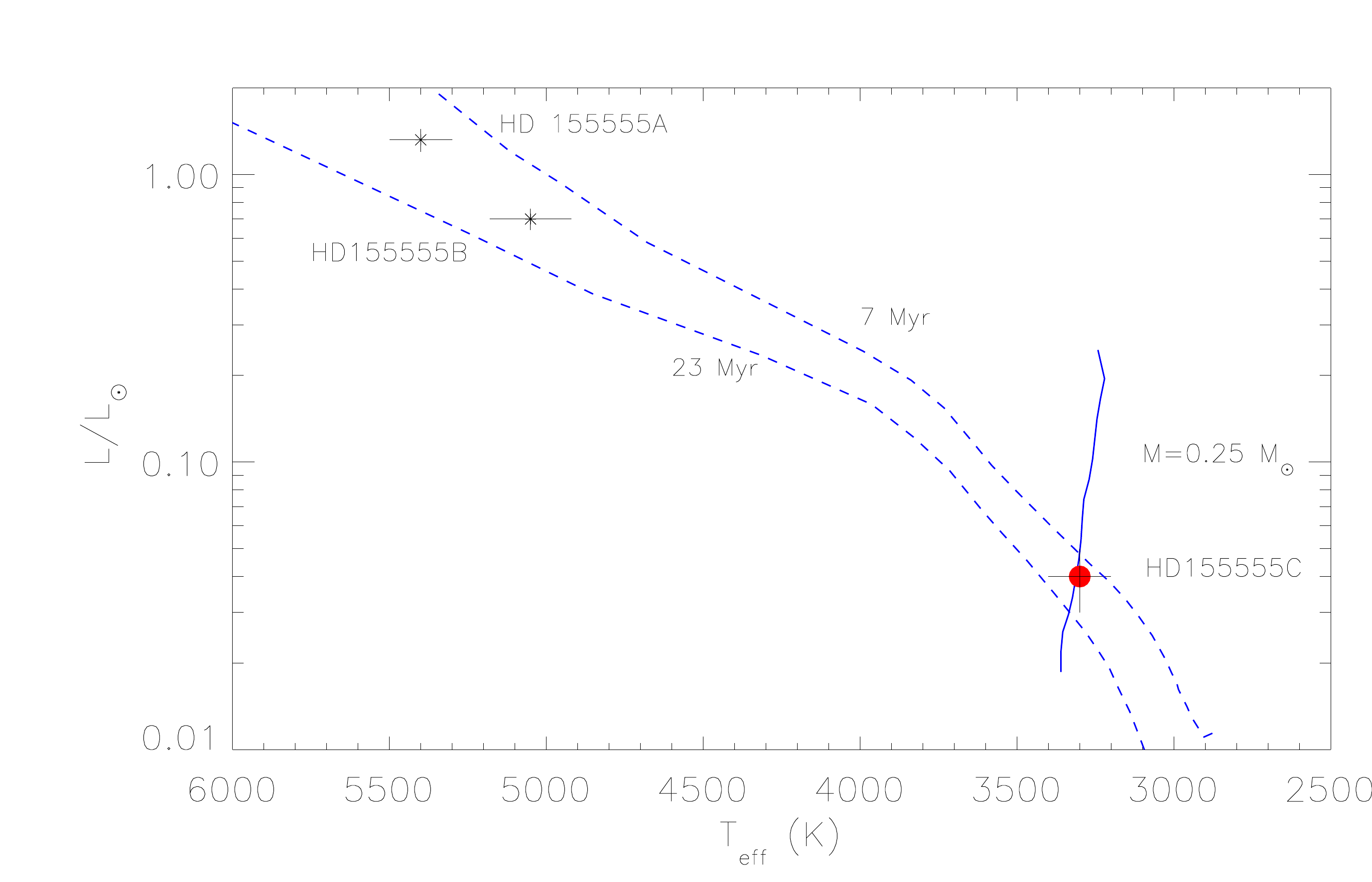} 
\end{minipage}
\caption{\label{hr}HR diagrams. Dashed lines are isochrones whereas the blue solid line is the  evolutionary track for a mass M=0.25 M$_\odot$. Models are from Baraffe et al. \cite{Baraffe98} (top panel), Siess et al. \cite{Siess00} (middle panel), and D'Antona \& Mazzitelli \cite{D'Antona97} (bottom panel).}
\vspace{0cm}
\end{figure}

We can use effective temperature and luminosity to derive the stellar radius and the inclination of the rotation axis.  We assume that HD155555AB and HD155555C are bound and, therefore, they have the same distance d = 31.4 pc as measured by Hipparcos for HD155555AB (Zuckerman et al. 2004). \rm Using the brightest observed magnitude V = 12.71 mag (Zuckerman et al. 2001), and the bolometric correction BC$_{\rm V}$ = $-$2.03\,mag  from Pecaut \& Mamajek \cite{Pecaut13}, we infer the bolometric magnitude M$_{\rm bol}$ = 8.19$\pm$0.05 mag, the luminosity L = 0.042$\pm$0.005\,L$_\odot$, and the stellar radius R = 0.65$\pm$0.05 R$_\odot$.  From these values and the stellar rotation period we infer the inclination of the stellar rotation axis  $i$ $\simeq$ 90$^\circ$ (adopting $<v \sin{i}>$ = 7.6 \,km\,s$^{-1}$). \\
In Fig.\,\ref{hr}, we compare T$_{\rm eff}$ and luminosity  of all components of \astrobj{HD\,155555} with a set of isochronones taken   from Baraffe et al. \cite{Baraffe98}, Siess et al. \cite{Siess00}, and D'Antona \& Mazzitelli  \cite{D'Antona97}, \rm for solar  metallicity to infer the age of each component and see if their coevalness hypothesis is sustainable. The physical parameters of \astrobj{HD\,155555A} and B are taken from Strassmeier \& Rice \cite{Strassmeier00}, with the only difference that luminosities are computed using the BC from Pecaut \& Mamajek \cite{Pecaut13}. 
 We see that the D'Antona \& Mazzitelli  \cite{D'Antona97} model produces the smallest age dispersion among all components with an estimated age of 15$\pm$7 Myr for the system and a  mass M=0.25 M$_\odot$ for \astrobj{HD\,155555C}. The Baraffe et al. \cite{Baraffe98} and Siess et al. \cite{Siess00} models produce results that emphasize a larger dispersion between the age of the SB2 components and the age of HD155555C. According to these models HD155555AB has an age A$_{\rm Baraffe}$ = 30$\pm$10 Myr and A$_{\rm Siess}$ = 24$\pm$8 My that are about a factor 2 older than the age of A$_{\rm Baraffe}$ = 13$\pm$7 Myr and  A$_{\rm Siess}$ = 12$\pm$4 Myr  of HD155555C.
 Despite these systematic differences between the Siess et al. \cite{Siess00} and Baraffe et al. \cite{Baraffe98} models and the D'Antona \& Mazzitelli \cite{D'Antona97} model, the uncertainties on the age are quite large and our results are consistent with the age range reported by Zuckermann et al. \cite{Zuckerman01} who found the system components to be fitted by a set of isochrones in the age range from 5 to 30 Myr. Thus, the coevalness hypothesis is still sustainable with an age of 20$\pm$15 Myr. Moreover, such an age is also consistent with the age of 21$\pm$9 Myr that Mentuch et al. \cite{Mentuch08} estimate for the $\beta$ Pic Association and based on the Li depletion. \rm
We note that the low Li content  (EW$_{\rm Li} < $ 20\,m\AA ) measured by Martin et al. \cite{Martin95} in the component C is consistent with the values measured on other Association members of similar spectral type.
On the contrary, the hotter components A and B are in the un-depleted part of the Li distribution, which prevents us from using the Li content to constrain their ages. 
\indent
 \begin{table*}
 \caption{\label{photometry} Photometry from the literature.}
 \begin{tabular}{cccccccccccc}
 \hline
U & B  & V & R & I & J & H & K & W1 & W2 & W3 & W4 \\
 (mag) & (mag) & (mag) & (mag) & (mag) & (mag)  & (mag) & (mag) & (mag) & (mag) & (mag) & (mag) \\
 \hline
 15.41 & 14.36 &  12.82 &   11.40 6 &  10.19  & 8.542 & 7.916 	 &	7.629 	&	7.528 	&	7.358 	&	7.204 	&	6.940 	\\
 \hline
  \end{tabular}
  \end{table*}\\

\begin{figure}
\begin{minipage}{10cm}
\includegraphics[width=90mm,height=120mm, angle=90,trim= 50 100 0 100]{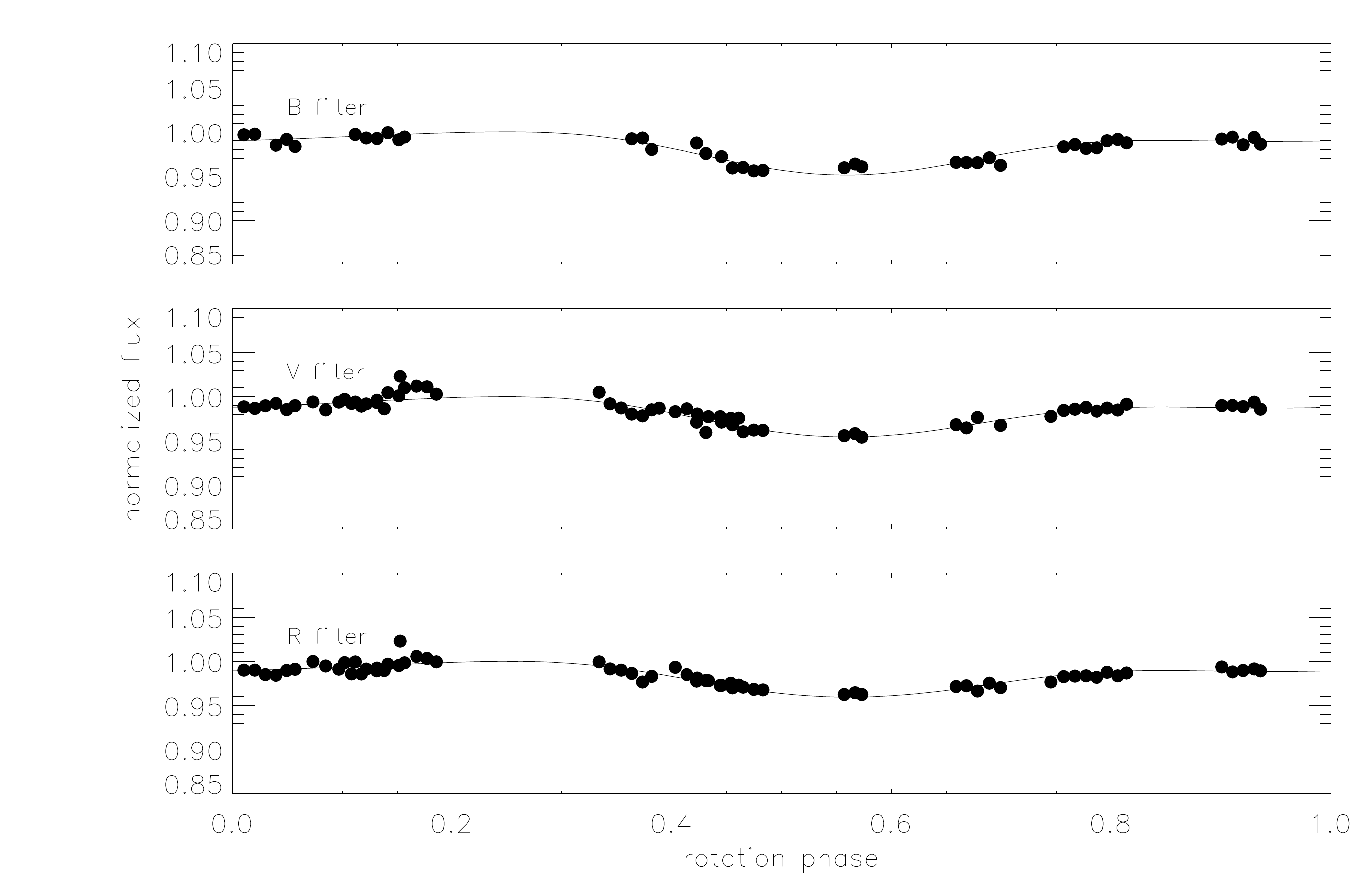}
\end{minipage}
\vspace{1cm}
\caption{\label{spotmodel} Spot model of HD155555C. Observed normalized flux (small bullets) in the B, V, and R filters versus rotation phase
 with overplotted the fit model (solid line). 
 }

\end{figure}

\begin{figure}
\begin{minipage}{10cm}
\includegraphics[width=110mm,height=100mm,angle=0,trim= 0 0 0 0]{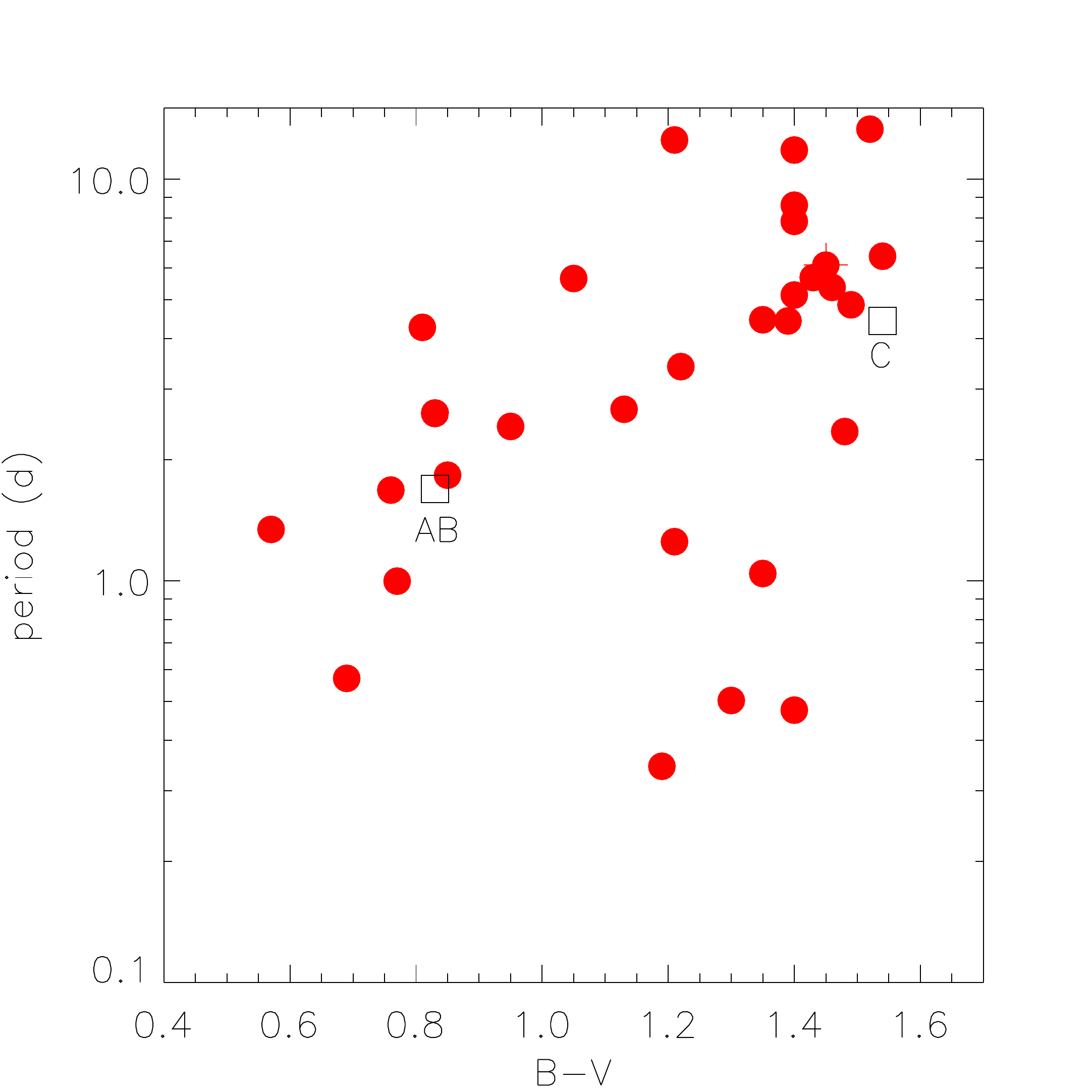} 
\end{minipage}
\caption{\label{period_distri}Rotation period distribution of $\beta$ Pic members from Messina et al. \cite{Messina11}. Squares denote the components of the \astrobj{HD\,155555} system.}
\vspace{0cm}
\end{figure}

\section{Spot model}
 To show that the presence of cool spots is the most typical  explanation of  the  light modulation observed in a low-mass star as HD155555C we performed a spot modeling of the light curves. \rm
We modeled the observed multi-band light curves using Binary Maker V 3.0 \cite{Bradstreet04}  to infer some information on the geometrical and physical properties of spots
on the photosphere of \astrobj{HD\,155555C}.  Binary Maker V 3.0 models are almost identical to those generated by Wilson-Devinney program \cite{Wilson71} and uses   Roche equipotentials to create star surfaces. \rm In our modeling the second component is essentially ``turned off" (i.e., assigned a near zero mass and luminosity) in order to model a single rotating star. The gravity-darkening coefficient has been assumed $\nu$ = 0.25 \cite{Kopal59}, and  limb-darkening coefficients from Claret et al.  \cite{Claret12} were also adopted. We adopted T$_{\rm eff}$ = 3300\,K, and inclination i = 90$^\circ$ as input parameters. To find a suitable spot configuration   we took into account that the light curve shape is asymmetric, the decreasing flux branch is shorter than the increasing flux branch. \rm This suggests the existence of  two major spot groups separated in longitude.  Unfortunately, we had no observations to be fit around the light curve maximum. \rm \\
Thanks to the dependence of the light curve amplitude on the photometric band wavelength, we could constrain the spot temperature contrast and better determine the area of the spots  responsible for the flux rotational modulation. \rm  The temperature contrast between spots and surrounding photosphere  was found  T$_{spot}$/T$_{phot}$ = 0.85, which is a value generally measured in dM3 spotted stars (see, e.g., Berdyugina 2005). The only free parameters in 
our model remained the  spot areas and the spot longitudes. The spots latitude cannot be constrained by photometry alone, especially in the case of an equator-on star as in our present case. \rm 
We found a satisfactory fit of V and R light curves using  two spots separated in longitude by  about 140$\pm$30$^{\circ}$ and with radius of about r = 25$\pm$5$^{\circ}$ each, which corresponds to about 17\% of the whole photosphere. Such a value refers to that component of spots unevenly distributed in longitude and, then, represents a lower value for the effective percentage of spotted surface. \rm In Fig.\,\ref{spotmodel}, we plot the normalized flux in the B, V, and R filters with overplotted the fits from our spot modeling.  The B model light curve is to some extent poorer with respect to the other fits, especially at the light minimum, where the observed light level is brighter than modeled. \rm


\section{Conclusion}
Our investigation allowed us to measure the rotation period P = 4.43d of the dM component \astrobj{HD\,155555C} in the triple system HD155555. Comparing all components luminosities and effective temperatures with PMS models, we find that the hypothesis that all three components are coeval is sustainable with an age of 20$\pm$15 Myr, which is consistent with literature values and with the age of the  $\beta$ Pic Association. Then, the difference of rotation periods can be explained considering the different nature of the components. The SB2 components that rotate with P = 1.687d have already reached a tidal synchronization, their  orbital \rm and rotation period being equal (Strassmeier et al. 2000; Cutispoto 1998). The single dM is still contracting and spinning up its rotation rate while approaching the zero-age main-sequence. The rotation periods of all components fit well  into the period distribution of other $\beta$ Pic members (see Fig.\,\ref{period_distri}).  From the spot modeling we infer a high level of photospheric activity, with spots having a temperature contrast of 0.85 with respect to the unspotted photosphere and a covering fraction  larger than about 17\% of the whole photosphere. This level is consistent with the high activity level at the higher atmospheric levels reported in the literature.

\section*{Acknowledgments}  
The Authors thank the anonymous Referees for their helpful comments that allowed to improve the quality of the manuscript. 
The extensive use of the SIMBAD and ADS databases, 
operated by the CDS center, (Strasbourg, France), is also gratefully acknowledged.

\end{document}